\documentstyle[twocolumn,aps,epsfig,prl,amsfonts,floats]{revtex}

\begin{document}

\title{Phase Transition in the Number Partioning Problem}
\draft
\author{Stephan~Mertens\thanks{email: Stephan.Mertens@Physik.Uni-Magdeburg.DE}}

\address{Universit\"at Magdeburg, Institut f\"ur Theoretische
Physik, Universit\"atsplatz 2, D-39106 Magdeburg, Germany
}

\date{\today}

\maketitle

\begin{abstract}
Number partitioning is an NP-complete problem of combinatorial
optimization. A statistical mechanics analysis reveals the existence of a phase
transition that separates the easy from the hard to solve instances
and that reflects the pseudo-polynomiality of number partitioning.
The phase diagram and the value of the typical ground state energy 
are calculated.
\end{abstract}


Computer science has recently discovered the notion of phase
transition in random combinatorial problems and its possible
connections with algorithmic complexity. In such a context,
statistical physics may provide an interesting perspective for
understanding problems in theoretical computer science. In this
letter we calculate the statistical mechanics of one of the core
problems in theoretical computer science.

The number partitioning problem is an easily formulated 
optimization problem: Given a set ${\cal A} = \{a_1,a_2,\ldots,a_N\}$ of
positive numbers,  find a partition, i.e.\ a subset 
${\cal A'}\subset {\cal A}$,
such that the residue
\begin{equation}
  \label{eq:costfunction}
  E = |\sum_{a_j\in {\cal A'}} a_j - \sum_{a_j\not\in {\cal A'}} a_j|
\end{equation}
is minimized.  
A partition with $E=0$ is called {\em perfect}.
The decision variant of the number
partitioning problem is to determine if there is a perfect partition
or not.

Number partitioning is of both theoretical and practical
importance. It is one of Garey and Johnson's six basic NP-complete
problems that lie at the heart of the theory of NP-completeness
\cite{garey:johnson:79}. Among
the many practical applications one finds multiprocessor scheduling
and the minimization of VLSI circuit size and delay.

A partition can be encoded by numbers $s_j=\pm 1$: $s_j=1$ if 
$a_j\in {\cal A'}$, $s_j=-1$ otherwise. The cost function then reads
\begin{equation}
  \label{eq:costfunction_spins}
  E = |\sum_{j=1}^Na_js_j|,
\end{equation}
and the minimum partition is equivalent to the ground state of the Hamiltonian
\begin{equation}
  \label{eq:hamiltonian}
  H = E^2 = \sum_{i,j=1}^N s_i\,a_ia_j\,s_j.
\end{equation}
This is an infinite range Ising spin glass with Mattis-like, antiferromagnetic 
couplings $J_{ij}=-a_ia_j$.  The thermodynamics of this model has been
investigated by Fu \cite{fu:89} and recently by Ferreira and Fontanari
\cite{ferreira:fontanari:98}. 

Fu claims that in the random number partitioning problem ``\ldots no
phase transition of any kind is found.'' \cite{fu:89}. If Fu were
right, number partitioning would be a notable exception to the
observation that many NP-complete problems do have a phase transition,
parameterized by a control parameter that separates the easy 
from the hard to solve instances \cite{cheeseman:etal:91}. 
For random, integer $a_i\in\{0,1,2,\ldots,A\}$,
Gent and Walsh \cite{gent:walsh:96} proposed
\begin{equation}
  \label{eq:kappa_GW}
  \tilde{\kappa} = \frac{\log_2A}N
\end{equation}
as a control parameter: they found numerically 
that one typically has $O(2^N)$ perfect partitions for $\tilde{\kappa} < \kappa_c$, 
whereas for $\tilde{\kappa} > \kappa_c$ the number of perfect partitions drops to zero.
The transition gets sharper with increasing $N$. Finite-size scaling
lead Gent and Walsh to $\kappa_c = 0.96$ for $N\to\infty$. 
This result contradicts Fu's claim, but
as we will see now, this type of phase transition can indeed be found
in the statistical mechanics of the number partitioning problem.

The canonical formalism of statistical mechanics requires the
calculation of the partition function
\begin{eqnarray}
  \label{eq:def_Z}
  Z &=& \sum_{\{S_i\}} e^{-\frac ET} = \sum_{\{S_i\}} \int_{-\infty}^{\infty}
  \!\!dx \,e^{-|x|}\, \delta(x - \frac{1}{T}\sum_{j=1}^N{a_js_j}) \nonumber\\
  &=& 2^N \int_{-\infty}^{\infty}\frac{d\hat{x}}{2\pi}\prod_{j=1}^N\cos(
  \frac{a_j}{T}\hat{x})\int_{-\infty}^{\infty}\!\!dx\,e^{-|x|+i\hat{x}x} \nonumber\\
    &=& 2^N \int_{-\frac{\pi}2}^{\frac{\pi}2}\!\!\frac{dy}\pi
  \prod_{j=1}^N \cos(\frac{a_j}T\tan(y))
\end{eqnarray}
where $T$ is the temperature. 
We write $Z$ as
\begin{equation}
  \label{eq:Z_average}
  Z = 2^N \int_{-\frac{\pi}2}^{\frac{\pi}2}\!\!\frac{dy}\pi e^{N\,G(y)}
\end{equation}
with
\begin{equation}
  \label{eq:G}
  G(y) = \frac1N\sum_{j=1}^N\ln{\cos(\frac{a_j}{T}\tan(y))}.
\end{equation}
At this point we could use the statistical independence of the $a_j$
and replace the sum by the average of $\ln\cos(\frac{a}{T}\tan(y))$
over $a$. This is usually done in Mattis-like spin glasses 
\cite{provost:vallee:83}, but we will proceed without this substitution
and calculate all thermodynamic quantities as functions of $\{a_j\}$.

For large $N$ the integral in Eq.~(\ref{eq:Z_average}) can be evaluated
using the saddlepoint technique. To find the saddlepoints of $G(y)$,
we will assume that $a$ can only take on values that are integer
multiples of a fixed number $\Delta a$. For integer distributions
$\Delta a = 1$, and for floatingpoint distributions $\Delta a$ is the
smallest number that can be represented with the available number of
bits. This assumptions leads to an infinite number of saddlepoints,
missed in \cite{ferreira:fontanari:98},
\begin{equation}
  \label{eq:saddlepoints}
  y_k = \arctan(\frac{\pi T}{\Delta a} k) \qquad k=0,\pm1,\pm2,\ldots.
\end{equation}
The resulting series of Gaussian integrals can be evaluated:
\begin{eqnarray}
  \label{eq:Z_annealed}
  Z &=& 2^N \sum_{k=0,\pm1,\ldots} \int_{-\infty}^{\infty}
    \!\!dy\,e^{-\frac{N}{2}G''(y_k)y^2} \nonumber\\
    &=& 2^N \sqrt{\frac{2 \Delta a^2}{\pi \sum_j a_j^2}}\,
  \coth\frac{\Delta a}{T}.
\end{eqnarray}
From that we get the average energy
\begin{equation}
  \label{eq:E_of_T}
  \frac{E}{T} = \frac{\Delta a}{T} \, \frac{\coth^2\frac{\Delta a}{T}-1}
  {\coth\frac{\Delta a}{T}}
\end{equation}
and the entropy
\begin{equation}
  \label{eq:entropy}
  S = N\ln2 - \frac12\ln\big(\frac{\pi \sum_j a_j^2}{2\Delta
    a^2}\big)
      + \tilde{S}(\frac{\Delta a}{2T}),
\end{equation}
where the thermal contribution reads
\begin{equation}
  \label{eq:thermal_entropy}
  \tilde{S}(\frac{\Delta a}{T}) = \ln\coth\frac{\Delta a}{T} 
  + \frac{\Delta a}{T} \, \frac{\coth^2\frac{\Delta a}{T}-1}
  {\coth\frac{\Delta a}{T}}.
\end{equation}
Note that for finte $\Delta a$, $\tilde S$ vanishes at zero temperature and increases
monotonically with $T$.
The entropy can be written as
\begin{equation}
  \label{eq:S_gent_walsh}
  S = N(\kappa_c-\kappa)\ln2 + \tilde{S}
\end{equation}
with 
\begin{equation}
  \label{eq:kappa_c}
  \kappa_c(N) = 1 - \frac{\ln\big(\frac{\pi}{6} N\big)}{N\,2\ln2}
\end{equation}
and 
\begin{equation}
  \label{eq:kappa}
  \kappa = \frac{\ln \frac{3}{\Delta a^2}\frac{1}{N}\sum_j a_j^2}{N\,2\ln2}.
\end{equation}
Note that $\kappa = \tilde{\kappa} + O(\frac{1}{NA})$ for the
distribution of the $a_i$'s 
considered by Gent and Walsh.

For $\kappa < \kappa_c$ the entropy is extensive even for 
$T=0$. According to Eq.~(\ref{eq:E_of_T}), the corresponding energy is
zero, hence we expect an {\em exponential number of perfect
  partitions}, in good 
agreement with the numerical results \cite{gent:walsh:96}.

For $\kappa > \kappa_c$ the zero temperature entropy seems to become negative. 
This would be wrong because the entropy must not
be smaller than $\ln 2$ for our discrete system. To see what is going
on here, note that $\kappa > \kappa_c$ means
\begin{equation}
  \label{eq:what_it_means}
  2^{-N} > \Delta a \sqrt{\frac{2}{\pi\sum_ja_j^2}}
\end{equation}
i.e.\ essentially $\Delta a = O(2^{-N})$. In this regime the
contributions of $\tilde S$ are $O(N)$ for any finite $T$,
\begin{equation}
  \label{eq:contrib_tilde_S}
  \tilde S(\frac{\Delta a}T) = \ln(\frac{T}{\Delta a}) + 1 +
  O(\frac{\Delta a^2}{T^2}),
\end{equation}
hence cannot be neglected.
Technically we deal with this contribution by
introducing an effective ``zero'' temperature $T_0$
below which the system can not be cooled. $T_0$ guarantees that
the contribution of $\tilde S$ remains $O(N)$. Its value
can be calculated from the lower bound of $S$:
\begin{eqnarray*}
  \ln 2 &=& N(\kappa_c-\kappa)\ln 2 + \tilde S(\frac{\Delta a}{T_0})\\
        &\approx& N(\kappa_c-\kappa)\ln 2 + \ln(\frac{T_0}{\Delta a}).
\end{eqnarray*}
From that we get
\begin{equation}
  \label{eq:T0}
  T_0 = 2\Delta a \, 2^{N(\kappa - \kappa_c)} = \sqrt{2\pi\sum\nolimits_ja_j^2}\,\,2^{-N}.
\end{equation}
For $\kappa > \kappa_c$ the ground state energy $E_0$ reads
\begin{equation}
  \label{eq:E_kappa}
  E_0 = T_0 = \sqrt{2\pi\sum\nolimits_ja_j^2}\,\,2^{-N}
\end{equation}
This equation specifies the rigorous result that the median value of
$E_0$ is $O(\sqrt{N}\,2^{-N})$ \cite{karmarkar:etal:86}.

\begin{figure}
\includegraphics[width=\columnwidth]{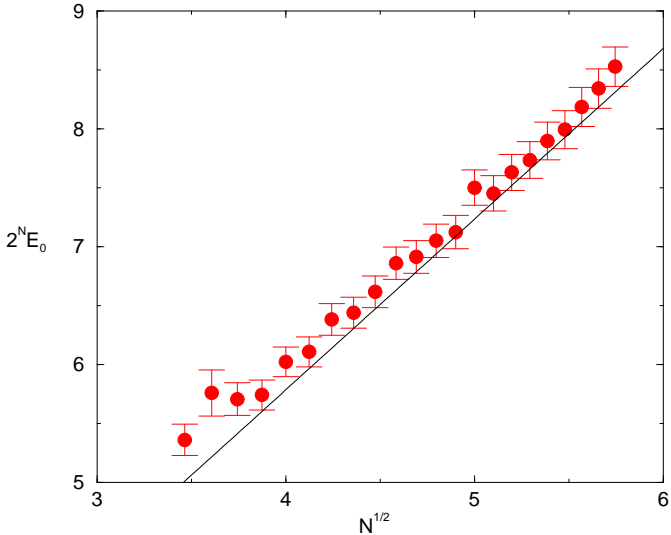}
\caption{
  Average minimum residue of the number partitioning problem
  with real numbers $0\leq a_i < 1$ compared to the analytical result 
  Eq.~(\ref{eq:E_kappa_cont}) (straight line).
    Each data point is the average over $10^4$ random samples.
\label{fig:eminuc_of_N}}
\end{figure}

To check Eq.~(\ref{eq:E_kappa}) we consider the continous variant of
number partitioning, where the $a_i$ are real
numbers, uniformely distributed in the interval $[0,1)$. 
In our formalism this means $\Delta a \to 0$ and
$\sum\nolimits_ja_j^2=N/3$.
We are in the $\kappa > \kappa_c$ regime and
Eq.~(\ref{eq:E_kappa}) becomes
\begin{equation}
  \label{eq:E_kappa_cont}
  E_0 = \sqrt{\frac23 \pi N}\,2^{-N} = 1.447\,\sqrt{N}\,2^{-N}
\end{equation}
In Fig.~(\ref{fig:eminuc_of_N}),  
Eq.~(\ref{eq:E_kappa_cont}) is compared to numerical data. The agreement
is convincing. The prefactor $\sqrt{\frac23\pi}$ fits
much better than the prefactor 
$\sqrt{\frac{\pi}{6e^2}} = 0.2662$,
reported in \cite{ferreira:fontanari:98}.

\begin{figure}
\includegraphics[width=\columnwidth]{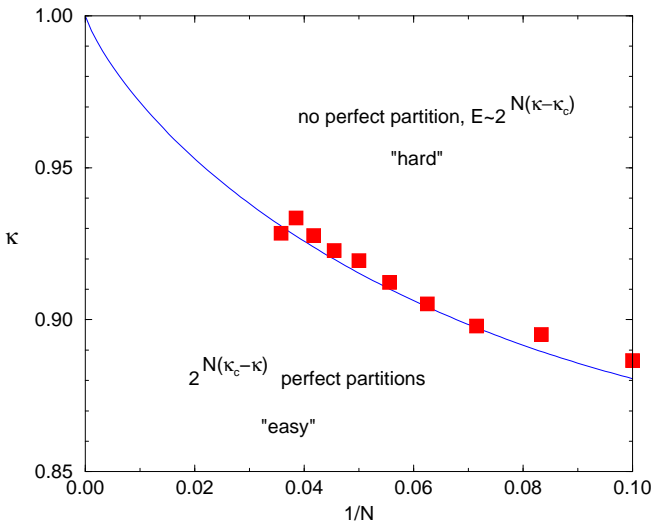}
\caption{
Phase diagram of the random number partioning problem. $N\kappa$
is essentialy the number of bits to encode the input numbers, see
Eq.~(\ref{eq:kappa}). The squares denote the phase boundary
found numerically. The solid line
is given by $\kappa_c$ from Eq.~(\ref{eq:kappa_c}).
For $\kappa<\kappa_c$, the zero temperature entropy is extensive and a
search algorithm typically finds quickly one of the $O(2^N)$ perfect
partitions. For $\kappa > \kappa_c$, no perfect partitions exist, 
and the optimization problem has a hard to find,
unique solution.
\label{fig:kappa_of_N}}
\end{figure}

To check whether $\kappa(N)$ is a control parameter with a phase
transition at $\kappa_c(N)$, we did numerical simulations. For fixed 
$N$ and $\kappa$ we calculated the
fraction of instances that have at least one perfect partition. In
accordance
with Gent and
Walsh \cite{gent:walsh:96} we find that this fraction is $1$ for small 
$\kappa$ and $0$ for larger $\kappa$. The transition from $1$ to $0$
is sharp. Fig.~(\ref{fig:kappa_of_N}) shows the numerically found
transition points for $10 \leq N \leq 28$ compared to 
$\kappa_c(N)$ from Eq.~(\ref{eq:kappa_c}). Again the agreement is
convincing. Note that $\kappa_c(N\to\infty)=1$. The asymptotic
estimate 0.96 given by Gent and Walsh is probably due to the 
rather small values $N\leq 30$ used in their simulations.

The two phases are very different with respect to the computational complexity 
of the corresponding instances. 
For $\kappa < \kappa_c$, a
search algorithm 
is likely to find one of the numerous perfect partitions in short
time, while the 
unique minimum partition for $\kappa > \kappa_c$ requires exponential
time to be found. 
This behavior can indeed be seen in numerical experiments
\cite{gent:walsh:96,korf:95,korf:98}. 
Refering to the solvability of the typical instance we call the two
phases ``easy'' and ``hard''. In a numerical investigation, the precision
of the numbers $a_i$ is fixed and $N$ is varied. For high precision and small
values of $N$, we are in the ``hard'' phase: no perfect partition exists, and
a search algorithm has to explore large parts of the configuration space.
If one increases $N$, the search space growth exponentially, and so does the
running time. On the other hand, increasing $N$ gets us closer to the phase 
boundary. Beyond this threshold, the number of perfect partitions
increases exponentially with $N$. A smart algorithm will 
try to find a perfect partition as 
early as possible. As can be seen from the experiments with the CKK
algorithm \cite{gent:walsh:96,korf:95,korf:98}, this may even lead to the effect 
that the running time now {\em decreases} with increasing $N$. As a function of $N$ the 
running time has a sharp maximum at the phase boundary: The hardest
problems are those close to the threshold.
A similar behavior has been found in other NP-complete problems like,
for example, the Satisfiability problem
\cite{hayes:97,monasson:zecchina:96,monasson:zecchina:97}. 

For bounded, integer values $0\leq a_i < A$ even the {\em worst case}
complexity of number partioning is polynomial in $A$ and $N$
\cite{garey:johnson:79}. 
This is no contradiction to the NP-completenes since a concise
encoding of an instance  
requires $N\log_2A$ bits, and $A$ is not bounded by a polynomial
function of $\log_2A$. 
Due to this property the number partitioning problem is called {\em
  pseudo polynomial} \cite{garey:johnson:79}. 
The exponential complexity of number partitioning relies on the
fact that extremely large (or precise) input numbers $a_i$ are
allowed.\footnote{
This distinguishes number partitioning
from many other NP-complete  
problems like for example, the travelling salesman problem, which
remains NP-hard even if the distances 
are restricted to take on the values $1$ and $2$. 
}
Pseudo polynomiality applies if the
number of bits to represent $a_i$ is fixed while $N$ increases.
This means, that $\kappa$ decreases, i.e.\ we get into the ``easy''
phase for large enough $N$. Hence the notion ``easy'' not only refers
to the {\em typical} (as shown here) but also to the {\em worst case} complexity.
This is a notable feature of the number partitioning problem:
The statistical mechanics results hold beyond the typical case for
which they are derived.

Looking at Eqs.~(\ref{eq:S_gent_walsh}, \ref{eq:E_kappa}) 
an interpretation of the parameter $\kappa_c$ suggests itself.
Let the $N$ numbers $a_i$ each be represented by
$N\kappa$ bits. Now consider the residue $E$ bitwise: About half of
all partitions will set the most significant bit of $E$ to zero. Among
those partitions, about one half will set the second most significant
bit to zero, too. Repeating this procedure we can set at most $N\kappa_c(N)$
bits to zero until running out of available partitions. If
$\kappa<\kappa_c$, we get a perfect partition before reaching this
point. The remaining set of available partitions has
$2^{N(\kappa_c-\kappa)}$ elements. This explains the zero temperature entropy, 
Eq.~(\ref{eq:S_gent_walsh}). For $\kappa>\kappa_c$, the
$N(\kappa>\kappa_c)$ least significant bits in $E$ can not be fixed
by the optimum partition, leading to Eq.~(\ref{eq:E_kappa}) for the
residue. 

\acknowledgements
The author appreciates stimulating discussions
with Andreas Engel and
R\'emi Monasson.

\bibliographystyle{prsty}
\bibliography{complexity,cs}

\end{document}